\documentclass[aps,twocolumn,showpacs,preprintnumbers,floatfix,amssymb]{revtex4} 

\usepackage{epsfig}
\usepackage{graphicx}

\newcommand{\beq}{\begin{equation}}
\newcommand{\eeq}{\end{equation}}
\newcommand{\beqa}{\begin{eqnarray}}
\newcommand{\eeqa}{\end{eqnarray}}

\newcommand{\cO}{\cal O}

\newbox\rotbox

\begin{document}

\preprint{
\hbox{ADP-03-123/T560}
}

\title{Comparison of $|Q|=1$ and $|Q|=2$ Gauge-Field Configurations\\
 on the Lattice Four-Torus}

\author{Sundance O. Bilson-Thompson}
\altaffiliation[Also at ]{School of Physics, Seoul National University, Seoul 151-747,
South Korea}
\email{sbilson@physics.adelaide.edu.au} 

\author{Derek B. Leinweber}
\email{dleinweb@physics.adelaide.edu.au}
\homepage{http://www.physics.adelaide.edu.au/theory/staff/leinweber/}

\author{Anthony G.\ Williams}
\email{anthony.williams@adelaide.edu.au}
\homepage{http://www.physics.adelaide.edu.au/theory/staff/williams.html}

\affiliation{CSSM Lattice Collaboration, Special Research Centre for the 
Subatomic Structure of Matter and The Department of Physics and 
Mathematical Physics, University of Adelaide, Adelaide SA 5005, Australia}

\author{Gerald V. Dunne}
\email{dunne@phys.uconn.edu}

\affiliation{Department of Physics, University of Connecticut, Storrs
  CT 06269, USA}

\date{\today}

\begin{abstract}
It is known that exactly self-dual gauge-field configurations with
topological charge $|Q|=1$ cannot exist on the untwisted continuum
4-torus.  We explore the manifestation of this remarkable fact on the
lattice 4-torus for $SU(3)$ using advanced techniques for controlling
lattice discretization errors, extending earlier work of De Forcrand
{\it et al} for $SU(2)$. We identify three distinct signals for the
instability of $|Q|=1$ configurations, and show that these signals
manifest themselves early in the cooling process, long before the
would-be instanton has shrunk to a size comparable to the lattice
discretization threshold. These signals do not appear for the
individual instantons which make up our $|Q|=2$ configurations. This
indicates that these signals reflect the truly global nature of the
instability, rather than the local discretization effects which cause
the eventual disappearance of the would-be single instanton.
Monte-Carlo generated $SU(3)$ gauge field configurations are cooled to
the self-dual limit using an ${\cal O}(a^4)$-improved gauge action
chosen to have small but positive ${\cal O}(a^6)$ errors.  This choice
prevents lattice discretization errors from destroying instantons
provided their size exceeds the dislocation threshold of the cooling
algorithm.  Lattice discretization errors are evaluated by comparing
the ${\cal O}(a^4)$-improved gauge-field action with an ${\cal
O}(a^4)$-improved action constructed from the square of an ${\cal
O}(a^4)$-improved lattice field-strength tensor, thus having different
${\cal O}(a^6)$ discretization errors.  The number of action-density
peaks, the instanton size and the topological charge of configurations
is monitored. We observe a fluctuation in the total topological charge
of $|Q|=1$ configurations, and demonstrate that the onset of this
unusual behavior corresponds with the disappearance of multiple-peaks
in the action density.  At the same time discretization errors are
minimal.
\end{abstract}


\maketitle


\section{Introduction}
\label{sec:intro}

Instantons are self-dual classical solutions of the Yang-Mills
equations which play several important roles in Quantum Chromodynamics
(QCD) \cite{jackiw,u1,shifman,schafer,pvb}.  Much is known about their
properties and their semiclassical consequences in flat
four-dimensional spacetime. On the other hand, at present the most
powerful nonperturbative approach to QCD is lattice QCD, where
spacetime is discretized. A discrete lattice with periodic boundary
conditions is in fact a four-toroidal mesh that approaches a continuum
4-torus in the continuum limit ($a \rightarrow 0$).  The properties of
instantons on a continuum four-torus ${\bf T}^4$ differ in interesting
ways from those of instantons on ${\bf R}^4$ \cite{torusreview}. In
particular, while simple explicit expressions for multi-instantons are
known on ${\bf R}^4$ (and on ${\bf S}^4$) via the ADHM construction
\cite{bpst,adhm}, no such concrete results are known on the continuum
four-torus. The only known explicit instanton solutions on ${\bf T}^4$
are quasi-abelian and of constant field strength
\cite{quasiabelian}. This lack of analytic information gives
additional motivation to the study of instantons in lattice QCD, where
instanton ({\it i.e.}\ self-dual) configurations can be obtained by
the technique of cooling, which has the goal of lowering the action
$S$ of the configuration without changing its topological charge
$Q$. The action is bounded below by the topological charge (in
suitable units), and a numerical instanton configuration is obtained
when this bound is saturated.

The {\it existence} on ${\bf T}^4$ of instantons of topological charge
$|Q|\geq 2$ was proved long ago by Taubes \cite{taubes}.  However, for
periodic boundary conditions it can be proved that $|Q|=1$ instantons
cannot exist on ${\bf T}^4$ \cite{braam}.  This is an elegant
corollary of a result known as the Nahm Transform \cite{nahm}, which
is an involution that maps an $SU(N)$ instanton of charge $Q$ on ${\bf
T}^4$ to an $SU(Q)$ instanton of charge $N$ on the dual torus ${\bf
\hat{T}}^4$. Since $U(1)$ does not support instantons, the Nahm
transform implies that $SU(N)$ cannot support $Q=1$ instantons.  The
unfortunate lack of exact torus instanton solutions makes explicit
study of Nahm's transform difficult, although it has been verified in
great detail \cite{qanahm} for the quasi-abelian constant field torus
instantons. Nahm's construction has been generalized
\cite{twistednahm} to incorporate twisted boundary conditions, which
describe additional non-abelian fluxes \cite{thoofttwists}. With such
twisted boundary conditions it is possible to have a $Q=1$ instanton
solution, and powerful lattice techniques have been developed to
numerically implement the Nahm transform including twisted boundary
conditions \cite{latticenahm}.  This is a {\it tour de force} of
lattice gauge theory, as it involves several highly nontrivial
numerical steps: cooling the configuration; finding zero modes of an
associated Weyl-Dirac operator; reconstructing the Nahm transformed
gauge field from the zero modes; and finally probing the self-duality
properties of the transformed configuration. The results provide
spectacular confirmation of the twisted Nahm transform construction
\cite{nahmsynthesis,pert}.

In this paper we address a somewhat different question, relevant for
instantons on an {\it untwisted} lattice.  We accurately cool
Monte-Carlo generated configurations in $SU(3)$ with a highly improved
action to produce highly self-dual configurations with a range of
topological charges. Our goal is to probe in a detailed manner the
difference(s) between the behavior under cooling of (untwisted)
configurations with topological charge $|Q|=1$ and topological charge
$|Q|=2$.  The possibility of observing the non-existence of self-dual
$|Q|=1$ configurations in numerical simulations of lattice gauge
theory on the 4-torus caught the attention of the lattice QCD
community some time ago \cite{deForcrand1997}.  There, configurations
with $|Q|=1$ were prepared by cooling Monte-Carlo generated $SU(2)$
configurations with the standard Wilson-Gauge action
\cite{deForcrand1995}.  The $|Q|=1$ configurations were then cooled
with improved actions toward the self-dual limit and were observed to
be unstable to improved cooling. The topological structures were
observed to shrink in size and eventually disappear from the lattice.
This was proposed as lattice evidence for the non-existence of $|Q|=1$
instantons on the four-torus \cite{deForcrand1997}.

Here, we extend this analysis in several ways, with the goal of
identifying and quantifying the impact of the Nahm transform corollary
on $|Q|=1$ configurations in $SU(3)$ gauge theory.  We find three
distinct signals of instability for the $|Q|=1$
configurations. Interestingly, these signals appear long before the
would-be instantons have shrunk to the point where they "fall through
the lattice". Furthermore, we show that these signals occur only for
the $|Q|=1$ configurations, and not for the individual separated lumps
that make up our $|Q|=2$ configurations, which shows that this is
really a global rather than a local effect, as expected for the
continuum Nahm transform corollary \cite{braam}.

A key part of our analysis is a precise monitoring and minimizing of
lattice discretization errors, which is crucial to interpreting the
results of cooling studies.  Errors in the discretization of the
action can destabilize instantons without regard to the topological
charge of the configuration.  Improved actions are formulated to
algebraically eliminate the leading finite lattice-spacing errors that
arise as a result of approximating continuous space-time by a discrete
mesh of points.  This enables simulations to more accurately approach
continuum behavior while still retaining a finite lattice spacing
\cite{symanzik}.  The use of such improved actions in cooling
algorithms has been shown to facilitate high-precision studies of the
properties of lattice gauge fields \cite{PerezandCo,deForcrand1995,
deForcrand1997,Bilson-Thompson:2002jk}.

A convincing demonstration of the instability of $|Q|=1$ lattice
instantons requires careful monitoring of the instanton size {\it and}
discretization errors.  These are particularly important, as any
instanton with a size smaller than the dislocation threshold of the
cooling algorithm is eliminated under improved cooling, again without
regard to the topological charge of the configuration.  The process is
signified by large discretization errors associated with the small
size of the object as it falls through the lattice.

This point was recognized as a caveat in the interpretation of the
results of Ref.~\cite{deForcrand1997}, as the size of the topological
structures in the configurations studied were initially very close to
the dislocation threshold of the cooling algorithm.  Thus, it was not
possible to determine whether the disappearance of the structure was
associated with the Nahm transform corollary, or simply the removal of
a dislocation under improved cooling.  Our results suggest that this
disappearance stage of the topological charge evolution under cooling
is in fact the final stage of an instability that manifests itself
much earlier in the cooling process, when the topological objects are
much larger than the dislocation threshold.

Following from the discussion above, we propose the following criteria
to identify convincingly the impact of the Nahm transform corollary:
\begin{enumerate}
\item The lattice-discretized cooling action must be accurate with
  remaining errors acting to stabilize topological structure.
\item The action and topological charge density of a $|Q|=1$
  configuration must be distributed over length scales much larger
  than the dislocation threshold of the improved cooling algorithm.
\item Evidence must be presented to confirm that discretization errors
  in the action are minimal where the Nahm transform corollary
  manifests itself.
\item Evidence that the action of the $|Q|=1$ configuration is
  approaching the self-dual limit of $8 \pi^2 / g^2$ must be provided.
\item Moreover, one must demonstrate that the distribution of action
  and topological charge distributions are approaching the classical
  form.
\end{enumerate}
We achieve each of these criteria, in the following manner:
\begin{enumerate}
\item We utilize a highly-improved action free from ${\cal O}(a^4)$
  errors and with slightly positive ${\cal O}(a^6)$ errors
  \cite{Bilson-Thompson:2002jk} to ensure the stability of instantons
  over several thousands of cooling sweeps.  This action is known as a
  three-loop improved action and appears better adapted to stabilizing
  topology than the five-loop improved action.
\item The size of the action and topological charge distributions are
  estimated by fitting the shape of the distribution surrounding the
  peak of the distribution to the classical single-instanton density
  profile.  The position and size, $\rho$, of the distributions are
  determined and compared with the dislocation threshold of the
  cooling action.
\item Remaining lattice discretization errors are evaluated by
  comparing the ${\cal O}(a^4)$-improved gauge-field action with an
  ${\cal O}(a^4)$-improved ``reconstructed action'' obtained from
  the square of an ${\cal O}(a^4)$-improved lattice field-strength
  tensor \cite{Bilson-Thompson:2002jk}, thus having different ${\cal
  O}(a^6)$ discretization errors.
\item Action and ``reconstructed action'' results will be reported in
  units of the single instanton action $S_0 = 8 \pi^2 / g^2$.
\item During the cooling process we also monitor the number of
  peaks in the action density.  Even as the action approaches the
  single instanton action, numerous peaks in the action density can be
  identified.  Their position and size are also monitored as a
  function of cooling sweep.
\end{enumerate}

This paper is set out as follows. In Section \ref{sec:LatActionandQ}
we briefly describe the highly-improved lattice discretization of the
continuum action, field-strength tensor, reconstructed action, and
topological charge operators.  In Section \ref{sec:Approach} we
outline the simulation techniques and parameters. Section
\ref{sec:Signature} examines the evolution of $|Q| = 1$ configurations
under improved cooling and puts the results in perspective through a
direct comparison with the behavior of $|Q| = 2$ configurations.  Our
conclusions are presented in Section \ref{sec:Conclusions}.

\section{Lattice Action and Topological Charge Operators}
\label{sec:LatActionandQ}

The lattice version of the Yang-Mills action was first proposed by
Wilson \cite{wilson}.  The action is calculated from the plaquette,
$W^{(1\times 1)}_{\mu \nu}$, a closed product of four link operators
incorporating the link $U_{\mu}$,
\begin{eqnarray}
S_{\mathrm{Wil}} & = & \beta \sum_{x} \sum_{\mu<\nu}
                \left(1-\frac{1}{N}Re\, {\mathrm{tr}}\,  W^{(1\times
                1)}_{\mu \nu}(x)\right)\, , \\ 
                 & = & \frac{1}{2} \int d^4x\, {\mathrm{tr}}\, F_{\mu \nu}^2 (x)
        +{\cal O}(a^2)\, ,
\label{eq:WilsonAction}
\end{eqnarray}
provided $\beta={2 N}/{g^2}$ for an $SU(N)$ field theory.  We will use
the notation $W^{(m \times n)}_{\mu\nu}$ to denote the closed loop
product (Wilson loop) in the $\mu-\nu$ plane with extent $m$ lattice
spacings in the $\mu$-direction and $n$ lattice spacings in the
$\nu$-direction.  Similarly, the lattice topological charge is
obtained by summing the charge density over each lattice site,
\beq
Q = {\displaystyle \sum_{x}} \frac{g^2}{32\pi^2}\epsilon_{\mu\nu\rho\sigma}
\ {\mathrm{tr}}\{F_{\mu\nu}(x)F_{\rho\sigma}(x)\}
\label{eq:LatQDens}
\eeq 
where $\mu, \nu, \rho, \sigma$ sum over the directions of the
lattice axes.

Since the action and topological charge constructed from different
Wilson loops have different ${\cal O}(a^2)$ and higher terms we may
cancel leading discretization errors by combining the contributions of
different loops. For example, de Forcrand {\em et al.}
\cite{deForcrand1995,deForcrand1997} have used tree-level improvement
to construct a lattice action which eliminates ${\cal O}(a^2)$ and
${\cal O}(a^4)$ errors, by using combinations of up to five Wilson
loop operators, which we denote $L_1,...,L_5$. In the case of
rectangular loops $m \neq n$ the contribution of the loops in each
direction is averaged, so that
\beqa
L_1 & \equiv & {W_{\mu\nu}^{(1 \times 1)}}\, , \nonumber \\
L_2 & \equiv & {W_{\mu\nu}^{(2 \times 2)}}\, , \nonumber \\
L_3 & \equiv & \frac{1}{2}\left\{ W_{\mu\nu}^{(2 \times 1)} +
W_{\mu\nu}^{(1 \times 2)}\right\}\, , 
            \nonumber \\
L_4 & \equiv & \frac{1}{2}\left\{ W_{\mu\nu}^{(3 \times 1)} +
W_{\mu\nu}^{(1 \times 3)}\right\}\, , 
            \nonumber \\
L_5 & \equiv & W_{\mu\nu}^{(3 \times 3)} \, . 
\eeqa
The improved action of de Forcrand {\em et al.} is a linear
combination of the Wilson actions calculated from $L_1,...,L_5$, each
divided by the relevant loop area squared (in units of $a^4$), and
respectively weighted by the constants
\begin{eqnarray}
c_1 & = & (19 - 55 c_5 ) / 9\, , \nonumber \\
c_2 & = & (1 - 64 c_5) / 9\, , \nonumber \\
c_3 & = & (640 c_5 - 64 ) / 45\, , \nonumber \\
c_4 & = & 1/5 - 2 c_5\, ,  
\end{eqnarray}
with $c_5$ as a free variable, i.e.,
\beq
S = c_1 S(L_1)+\frac{c_2}{16}S(L_2)+\frac{c_3}{4}S(L_3)+\frac{c_4}{9}S(L_4)+\frac{c_5}{81}S(L_5).
\eeq
Selecting appropriate values for $c_5$ enables one to set the
contribution of certain loops to zero, creating so-called 3-loop
($c_5=1/10$), or 4-loop ($c_5=0$) improved actions. Other values of
$c_5$ lead to 5-loop improved actions.  Following
Ref.~\cite{deForcrand1995,deForcrand1997}, we consider $c_5=1/20$
(midway between the 3-loop and 4-loop values) to define the
``standard'' 5-loop action.

All of these actions have the same level of improvement, but in
general they will have different ${\cal O}(a^6)$ errors.  We refer to
the Wilson action, Eq.~(\ref{eq:WilsonAction}), constructed from the
plaquette as a 1-loop action. A 2-loop action may be constructed
\cite{LePage} from $\frac{5}{3}L_1 - \frac{1}{6}L_3$.  These
definitions of the variously-improved actions are used in our cooling
algorithm.

We construct a highly-improved lattice field-strength tensor,
analogously to the construction of the cooling action.  The relations
\begin{eqnarray}
W^{(1 \times 1)}_{\mu \nu}     & = & 
        \left\{1+ig\oint A.dx -\frac{g^2}{2}(\oint A.dx)^2
         + {\cal O}(g^3) \right\}\, , 
                                                        \nonumber \\ 
W^{(1 \times 1){\dagger}}_{\mu \nu} & = & 
        \left\{1-ig\oint A.dx -\frac{g^2}{2}(\oint A.dx)^2
         + {\cal O}(g^3) \right\}\, ,
                                                        \nonumber 
\end{eqnarray}
indicate the field-strength tensor may be accessed via
\begin{eqnarray}
\frac{-i}{2} && \hspace{-0.4cm} \left(W^{(1 \times 1)}_{\mu \nu} - 
                     W^{(1 \times 1){\dagger}} 
              - \frac{1}{3} {\mathrm{tr}} \left (W^{(1 \times 1)}_{\mu \nu} - 
                            W^{(1 \times 1){\dagger}}_{\mu\nu}\right )\right)
\nonumber \\
         &&= g\oint A.dx + {\cal O}(g^3) \nonumber \\
         &&= ga^2 F_{\mu \nu}(x_0) + {\cal O}(ga^4) + {\cal O}(g^3a^4)
\, ,
\label{eq:ExtractIntegral}
\end{eqnarray}
where we have subtracted one-third of the trace to enforce the
traceless aspect of the Gell-Mann matrices.  As with the improved
action, the improved field-strength tensor is constructed from a
combination of Wilson loops.  Our improved field-strength tensor is
given by
\begin{eqnarray}
g\, F^{\mathrm{Imp}}_{\mu\nu} &=& \left[ 
                 k_1 C^{(1,1)}_{\mu \nu} + 
                 k_2 C^{(2,2)}_{\mu \nu} +
                 k_3 C^{(1,2)}_{\mu \nu} + \right . \nonumber \\
&& \quad \left . k_4 C^{(1,3)}_{\mu \nu} +
                 k_5 C^{(3,3)}_{\mu \nu} \right] \, ,
\end{eqnarray}
where $C^{(m,n)}_{\mu \nu}$ are the clover averages of $m \times n$
and $n \times m$ path-ordered link products
\cite{Bilson-Thompson:2002jk}.  The improvement constants take the
values
\begin{eqnarray}
k_1 & = & 19/9 - 55 k_5\, , \nonumber \\
k_2 & = & 1/36 - 16 k_5\, , \nonumber \\
k_3 & = & 64 k_5 - 32/45\, , \nonumber \\
k_4 & = & 1/15 - 6 k_5\, , \nonumber 
\end{eqnarray}
and in this case $k_5$ is the tunable free parameter.
$F^{\mathrm{Imp}}_{\mu\nu}$ may be inserted directly into
Eq.~(\ref{eq:LatQDens}) to create an improved topological charge
operator, or into the equation
\cite{Bilson-Thompson:2002jk,correction}
\beq
S_R = \beta \sum_{x} \sum_{\mu,\nu} \frac{1}{12}\, \mathrm{tr} \,
      \left [ g^2 \, F_{\mu \nu}^2(x) \right ] \, . 
\eeq
to create a ``reconstructed'' improved action, $S_R$.  Both improved
operators will have errors of ${\cal O}(a^6)$.  Errors associated with
$g$ are rapidly suppressed in the process of cooling.  Appropriate
choices of $k_5$ enable us to create 3-loop or 4-loop improved
versions of the field-strength.  Our 5-loop improved field-strength
tensor is defined with $k_5=1/180$, midway between the 3-loop and
4-loop values of $k_5 = 1/90$ and $0$ respectively
\cite{Bilson-Thompson:2002jk}.  While the 5-loop improved
field-strength tensor is marginally more accurate, it is
computationally expensive.  We adopt the 3-loop improved
field-strength tensor to construct the ${\cal O}(a^4)$-improved
topological charge and reconstructed action operators.

Tadpole corrections to the improvement coefficients \cite{LePage} are
included and updated after every cooling sweep.  Our experience
\cite{Bilson-Thompson:2002jk} is that tadpole improvement factors are
beneficial in the early stages of cooling and remain beneficial even
as $u_0 \rightarrow 1$.

Previous investigations \cite{Bilson-Thompson:2002jk} have indicated
the 3-loop and 5-loop improved actions are similarly accurate in
reproducing the classical action of approximately self-dual
configurations.  In some cases, the 5-loop action would underestimate
the classical action.  Such errors open the possibility of
destabilizing topological structures over several thousand cooling
sweeps, regardless of the topological charge of the configuration.
For the configurations investigated in
Ref.~\cite{Bilson-Thompson:2002jk} the errors in the 3-loop improved
action were consistently positive; thus acting to stabilize topology.
In the following, we adopt the 3-loop improved action.

\section{Lattice Approach}
\label{sec:Approach}

To investigate the stability of self-dual gauge field configurations
on the (untwisted) 4-torus, we construct an ensemble of field
configurations using the Cabibbo-Marinari \cite{CabMar}
pseudo-heatbath algorithm with three diagonal $SU(2)$ subgroups looped
over twice.  We thermalize for 5000 sweeps with an ${\cal
O}(a^2)$-improved action from a cold start (all links set to the
identity) and select configurations every 500 sweeps thereafter
\cite{Bonnet:2001rc}.  Configurations are numbered consecutively in
the order that they are produced in the Markov-chain process.  Hence
configuration 1 was saved after 5000 thermalization sweeps from a cold
start, configuration 2 was saved 500 sweeps after configuration 1, and
so on.  Our results are generated on a $12^3 \times 24$ periodic
lattice at $\beta = 4.60$, with a lattice spacing of $a=0.122(2)$ fm
determined by a string tension analysis incorporating the lattice
coulomb term.  Cooling is performed with a 3-loop improved action with
tadpole improvement (using the plaquette definition of the mean link)
and the topological charge is assessed with the 3-loop improved
operator.  Links which may be updated in parallel are identified and
partitioned \cite{Bonnet:2000db}.  Updating the partitioned links in
parallel minimizes the drift of objects in the cooled configurations.

As a thermalized configuration is cooled, over the course of many
hundreds of sweeps the action $S$ of the configuration monotonically
decreases.  This occurs because the cooling algorithm smooths out
short-range fluctuations in the field.  As the high-frequency
components of the field are suppressed the underlying semi-classical
structure of the field is revealed. If cooling proceeds for long
enough the configurations will become self-dual consisting only of 
instantons or anti-instantons.

The action, $S$, is bounded below by the magnitude of the topological
charge, $Q$, i.e.,
\beq
S \geq S_0\, |Q| = \frac{8 \pi^2}{g^2}\, |Q| \, , 
\label{eq:SQinequality}
\eeq
where in the continuum $Q$ assumes an integer value
\cite{shifman,schafer}.  The quantity $S_0= \frac{8 \pi^2}{g^2}$ is
the action associated with a single instanton and is independent of
the instanton size.

As cooling is applied uniformly over the 4-volume of the torus,
configurations become locally self-dual.  In the infinite volume limit
where well-separated instantons and anti-instantons can simultaneously
be present, the total action and topological charge will approach
discrete values, satisfying the relations $S/S_0=n_I+n_A$ and
$Q=n_I-n_A$, where $n_I$ and $n_A$ are the number of instantons and
anti-instantons present respectively.  On a finite volume, one might
expect to see some plateauing in the action density as a function of
cooling sweep, when these conditions are satisfied.  However, further
cooling will lead to the annihilation of instanton--anti-instanton
pairs in the finite volume, until only instantons or only
anti-instantons remain.  The configuration will then continue to cool
until it achieves complete global self-duality, i.e., $S/S_0=|Q|$.

As discussed in the Introduction, self-dual configurations with
$|Q|=1$ ({\it i.e.}\ a single instanton or a single anti-instanton) do
not exist on the untwisted continuum four-torus \cite{braam}. Note, of
course, that configurations with topological charge $|Q|=1$ do exist;
they just cannot be made self-dual. In other words, the lower bound
$S\geq S_0$ cannot be exactly saturated. We therefore expect to see
different behavior of $|Q|=1$ and $|Q|=2$ lattice configurations under
cooling. The next section is devoted to a study of signatures of this
different behavior.

\section{Cooling comparison for $Q=1$ and $Q=2$ configurations}
\label{sec:Signature}

\subsection{Action and Charge Evolution}
\label{cresting}

Throughout the discussion of the results we will use the following
notation: the cooling action will be denoted by $S$, the reconstructed
action by $S_R$ and the topological charge by $Q$. The type of
improvement scheme used will be denoted by a number in
parentheses. Hence the 2-loop improved cooling action, constructed
from $L_1$ and $L_3$ is written as $S(2)$, our topological charge
calculated from a 3-loop improved field-strength tensor is written as
$Q(3)$, our 3-loop reconstructed action (calculated from a 3-loop
improved field-strength tensor) is written as $S_R(3)$, and so on.
The action reported throughout the cooling process will typically be
divided by $S_0$ to facilitate the comparison with $|Q|$. Hence the
curves in figures labeled as $S(n)$ and $S_R(n)$ actually represent
the normalized action values $S(n)/S_0$ and $S_R(n)/S_0$ respectively.

\subsubsection{$|Q|=1$ Configurations}

\begin{figure}[t]
\begin{center}
\epsfig{figure=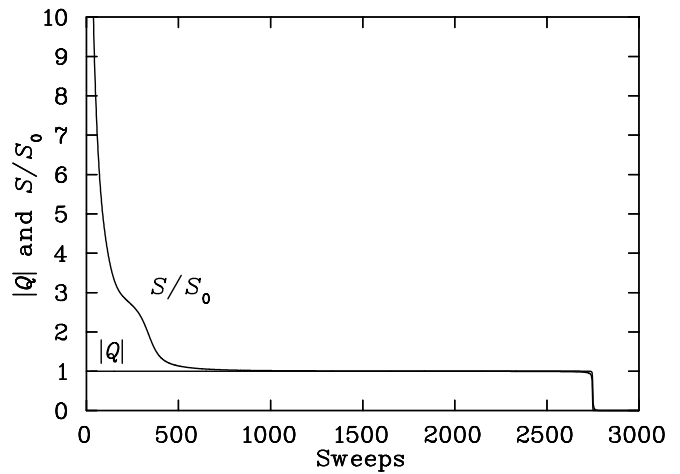,angle=90,width=\hsize}
\end{center}
\caption{ Topological charge, $Q(3)$, and normalized action,
$S(3)/S_0$, for configuration 64.  Notation is described in the text.
In particular, $Q$ and $S$ are assessed here with the 3-loop improved
topological charge and action operators.  The action evolution
flattens around 300 sweeps characteristic of a well separated
instanton anti-instanton pair.  Annihilation of the instanton
anti-instanton pair follows as signified by the reduction of the
action by two units, while the topological charge remains unchanged.
The nearly self-dual single instanton configuration destabilizes
around sweep 2700.  }
\label{fig:cfg64zout}
\end{figure}

In Fig.~\ref{fig:cfg64zout} we show $S(3)/S_0$ and $|Q(3)|$ for
configuration 64.  After a small number of cooling sweeps this
configuration rapidly settles down to a topological charge of one.
Hence we expect that under continued cooling with negligible
discretization errors, it should never reach true self-duality.  We do
observe a long plateau where the configuration appears to closely
approach self-duality.  However, around cooling sweep 2700 it
eventually destabilizes and collapses to triviality.  This behavior is
consistent with earlier results \cite{deForcrand1997} for $SU(2)$.  We
will argue in the following that this final destabilization of the
instanton is in fact the final stage of a series of evolution stages,
with this final disappearance of the would-be instanton being due to
its small size relative to the dislocation threshold of the cooling
algorithm.

\begin{figure}[t]
\begin{center}
\epsfig{figure=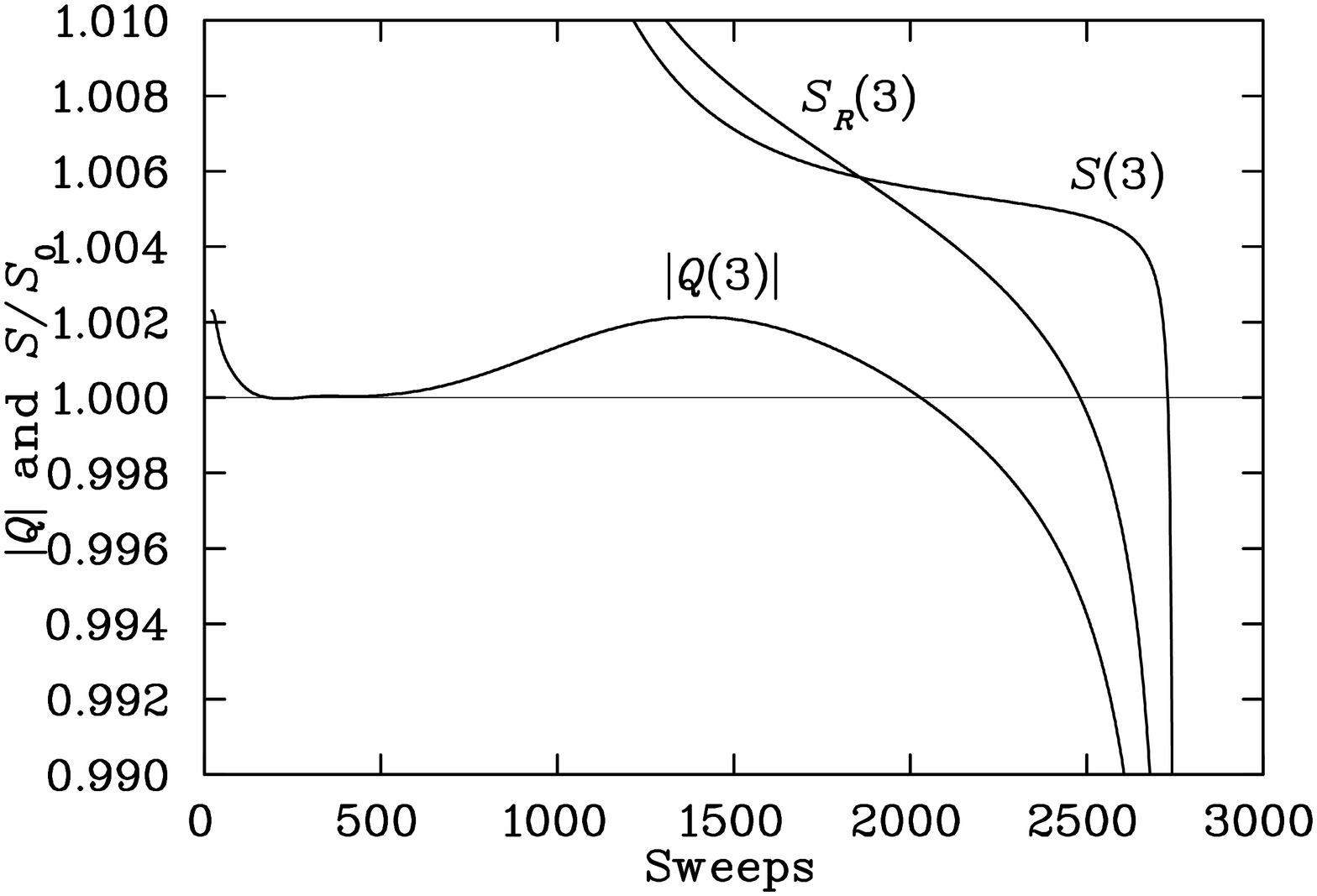,angle=90,width=\hsize}
\end{center}
\caption{Plot showing the fine detail of the $|Q|=1$ configuration
  illustrated in Fig.~\protect\ref{fig:cfg64zout}; configuration~64.
  The three-loop cooling action, $S(3)$, three-loop reconstructed
  action, $S_R(3)$, and the three-loop topological charge, $Q(3)$ are
  plotted.  A very high level of accuracy is obtained for the
  topological charge between sweeps 200--400.  A ``cresting behavior''
  appears as $S/S_0$ drops from $\sim 3$ toward 1.}
\label{fig:cfg64zin}

\begin{center}
\epsfig{figure=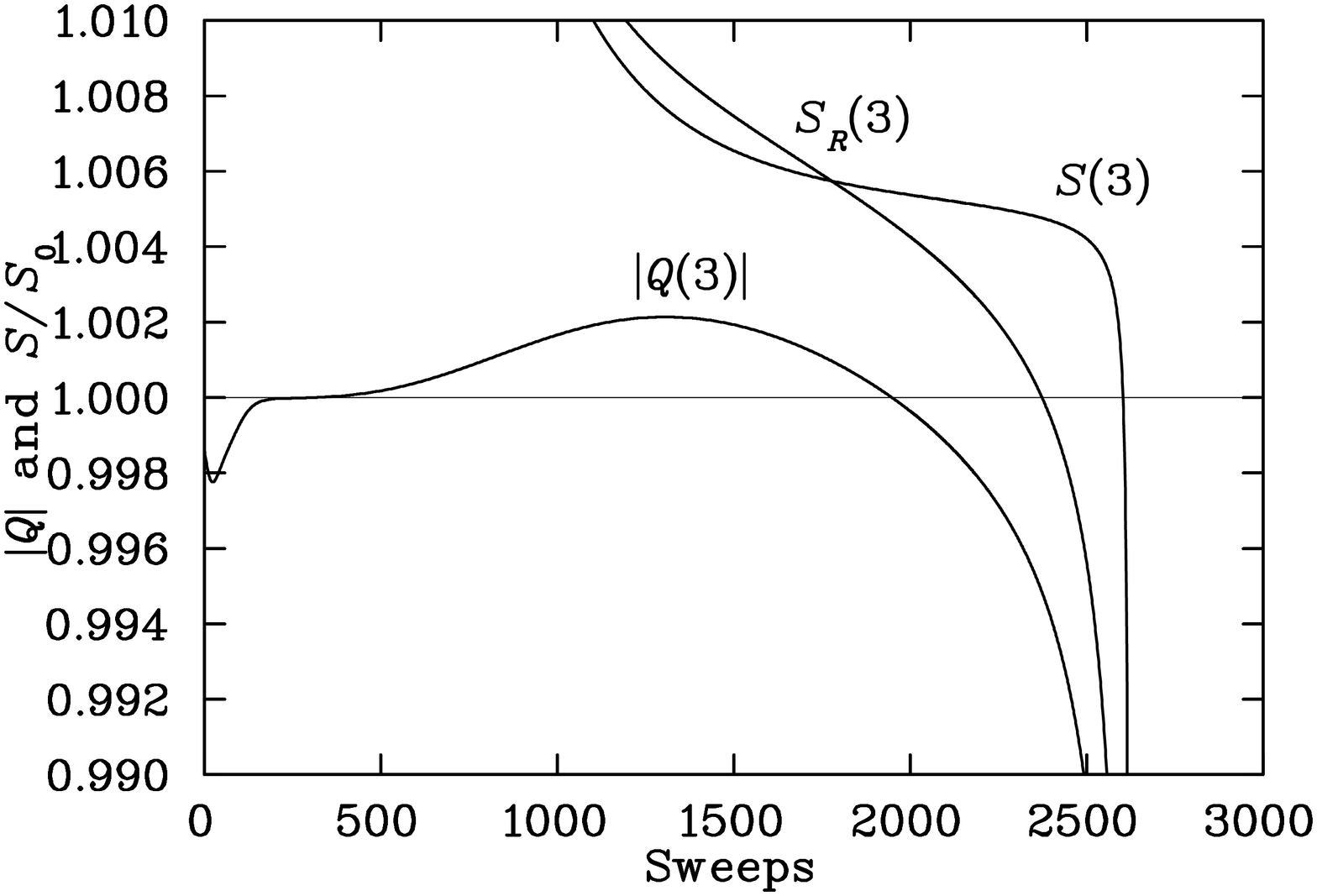,angle=90,width=\hsize}
\end{center}
\caption{The fine detail of another $|Q|=1$ configuration;
  configuration 11.  Beyond 200 sweeps, the behavior is nearly
  identical to that illustrated in Fig.~\protect\ref{fig:cfg64zin}.  }
\label{fig:cfg11zin}
\end{figure}

Let us now examine the same configuration with an enhanced vertical
scale as in Fig.~\ref{fig:cfg64zin}.  The configuration fails to
achieve self-duality before it destabilizes and becomes trivial.
However, a very high level of accuracy is obtained for the topological
charge, $Q$, between sweeps 200--400.  This is the region in which the
action evolution flattens with $S/S_0 \simeq 3$ signaling approximate
local self-duality.  This region provides a benchmark for the level of
accuracy that may be achieved through the use of highly improved
operators.

For sweeps 400-1400 we see the value of $Q$ rise away from the integer
value of one.  This ``cresting behavior'' commences at the same time
that the action leaves the regime of $S/S_0 \simeq 3$ and approaches
the classical instanton action.  We note that it is well separated from
the ultimate collapse of the configuration.

The two 3-loop improved actions ($S(3)$ and $S_R(3)$) remain greater
than $S_0 |Q|$ as required.  The two actions deviate significantly
from one another when the configuration collapses to triviality, which
is indicative of large discretization errors.  Thus, this
destabilization, at this stage of the cooling process, should not be
attributed to the Nahm transform corollary.  In contrast, the two
action measures agree relatively well even at the peak in the
topological charge cresting.  Note that $S(3)/S_0 > 1$, so that errors
should act to stabilize the configuration.

To illustrate the reproducible nature of these results we also present
results in Fig.~\ref{fig:cfg11zin} for configuration~11; another
$|Q|=1$ configuration.  Beyond 200 sweeps, the behavior is nearly
identical. Beyond approximately 1500 cooling sweeps we see a crossing
followed by a large divergence of the action values and a sudden drop
in the topological charge.  This indicates the onset of severe
discretization errors, again suggesting a small size for the
nontrivial topological structure, the ``would-be'' instanton.

Based on these results, we propose that the cresting of the
topological charge early in the cooling process, and associated with
the departure of $S(3)/S_0$ from the regime of 3 toward 1, is a
distinct signature of the ultimate instability of the single instanton
on the lattice.  The eventual disappearance of the would-be instanton
happens much later in the cooling process, and appears to rather be a
consequence of large discretization errors in that region.  In the
next sub-section, we contrast this directly with the behavior of
$|Q|=2$ configurations under cooling.

\subsubsection{$|Q|=2$ Configurations}

\begin{figure}[t]
\begin{center}
\epsfig{figure=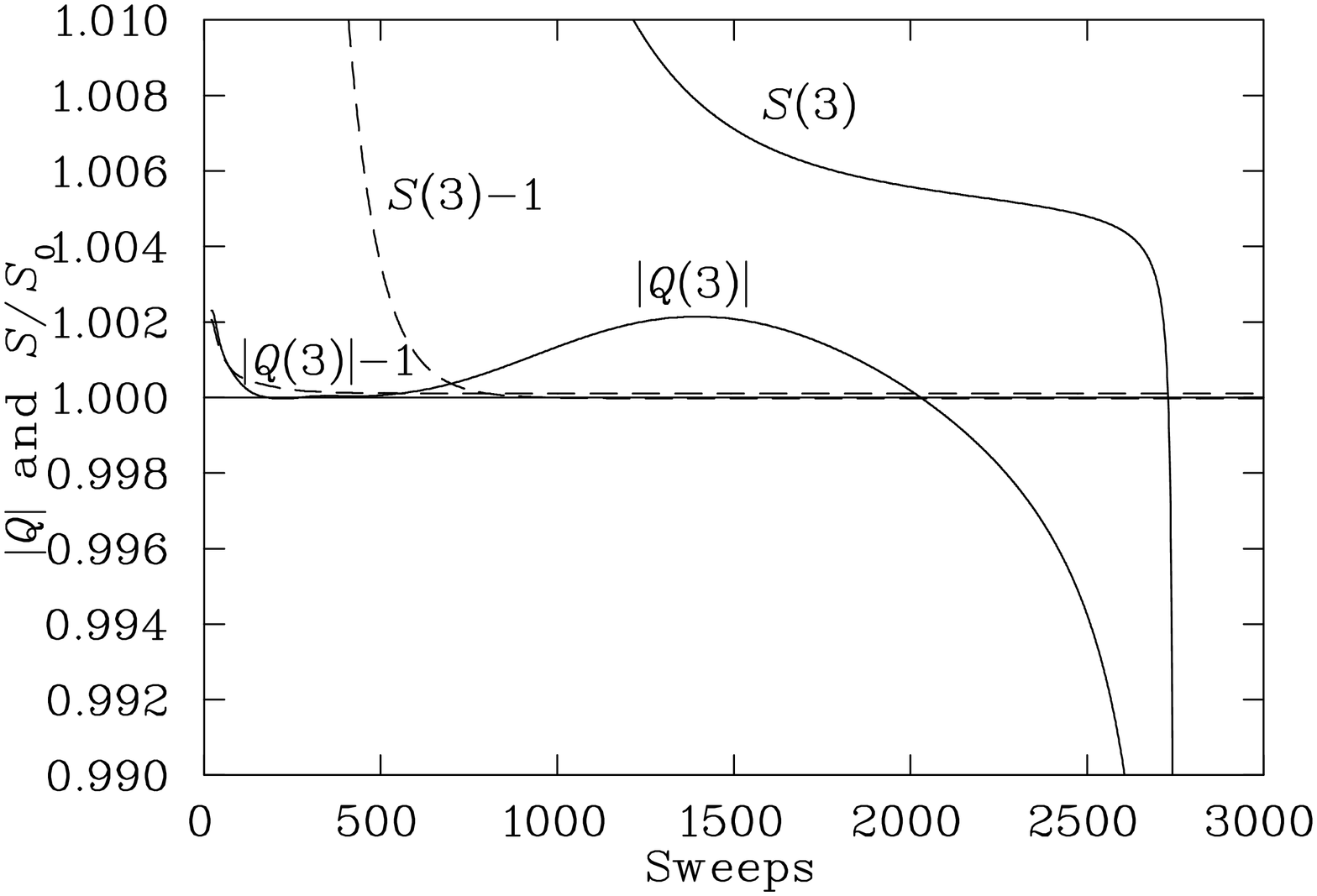,angle=90,width=\hsize}
\end{center}
\caption{Comparison of the $|Q|=1$ configuration (configuration 64,
solid line) with a $|Q|=2$ configuration (configuration 90, dashed
line).  For the two-instanton configuration both $S(3)/S_0$ and $|Q|$
have been reduced by unity so that the curves may be directly compared
with the $|Q|=1$ configuration.  The accurate behavior of the
two-instanton configuration contrasts the single-instanton case.  }
\label{fig:cfg90and64zin}

\begin{center}
\epsfig{figure=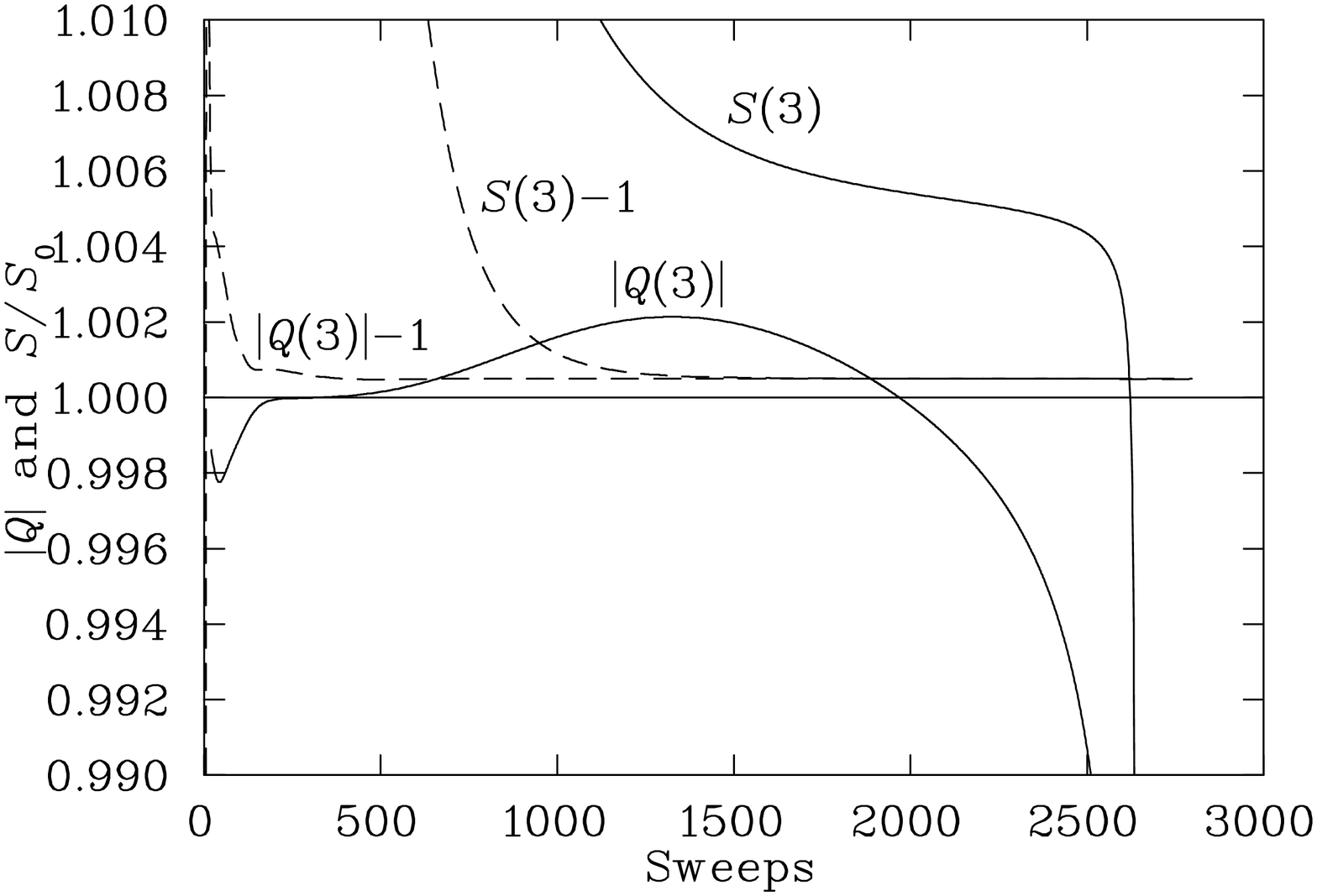,angle=90,width=\hsize}
\end{center}
\caption{A similar comparison of a $|Q|=1$ configuration
(configuration 11, solid line) with a $|Q|=2$ configuration
(configuration 27, dashed line).  As in
Fig.~\protect\ref{fig:cfg90and64zin}, $S(3)/S_0$ and $|Q|$ for
the two-instanton configuration have been reduced by unity to
allow a direct comparison of the curves with the $|Q|=1$
configuration.}
\label{fig:cfg11and27zin}
\end{figure}

To show that this cresting behavior is specific to $|Q|=1$
configurations, we compare one and two instanton configurations in
Figs.~\ref{fig:cfg90and64zin} and \ref{fig:cfg11and27zin}.  In these
figures we consider the three-loop quantities $S(3)/S_0$ and $|Q(3)|$.
For the two-instanton case, we plot $(S(3)/S_0) - 1$ and $|Q(3)| - 1$,
so that the curves for the $|Q|=1$ and $|Q|=2$ cases may be directly
compared.  The level of self-duality achieved by the two-instanton
configurations is far superior to that of the single-instanton
configurations.  The cresting behavior observed in the topological
charge of the $|Q|=1$ configurations is not seen at all for the
$|Q|=2$ configurations.

We stress that the behavior of the $|Q|=1$ configurations differ
significantly from that of the $|Q|=2$ configurations long before the
$|Q|=1$ configurations collapse to triviality.  Indeed, this deviation
becomes manifest relatively early in the cooling process.  For
configuration 64 of Figs.~\ref{fig:cfg64zout} and \ref{fig:cfg64zin}
the cresting behavior begins as the number of locally self-dual
objects on the lattice drops from three to one.

To confirm the consistency of this picture, it is important to know
the size of these objects in order to evaluate whether the dislocation
threshold of the cooling algorithm is playing a role in the evolution
of the configurations, and at what stage of the cooling process this
is important.  This is the subject of the next section.

\subsection{Instanton size evolution}
\label{sec:SizeEvolu}

To investigate further the behavior of the topological structure of
these $|Q|=1$ and $|Q|=2$ configurations under cooling, we implement
an algorithm which identifies local peaks in the action and
topological charge densities.  Peaks in the action density are
identified by finding a point at the center of a $3^4$ hypercube whose
action density exceeds that of the neighboring 80 points of the
hypercube.  The algorithm may also be applied to the topological
charge density in two steps, reversing the sign of the topological
charge density to convert valleys to peaks.

The algorithm then fits the structure of the gauge fields around these
peaks to the classical instanton form for the action density
\begin{equation}
S(x) = \xi \, \frac{6}{\pi^2} \, \frac{\rho^4}
  {((x-x_0)^2 + \rho^2))^4}\, ,
\label{eq:instantonform}
\end{equation} 
generalized by the inclusion of $\xi$ to allow for an overall
normalization different from 1 due to periodic images of the action
density \cite{Kusterer:2001vk}.  For an instanton in infinite volume
on ${\bf R}^4$, $\xi=1$.  The fit parameters returned by this
algorithm are the coordinates of the center of each instanton-like
peak, $x_0$, the instanton-size parameter for each peak, $\rho$, and
the overall scale factor for the peak, $\xi$.

It should be noted that not every peak found by this algorithm is an
instanton, particularly during the early stages of the cooling
process.  However, we certainly anticipate that the peaks which
survive under cooling will approach the form of instantons or
anti-instantons as the cooling proceeds.  Peaks which do not
correspond to instanton-like structures will disappear in the early
stages of cooling.  Of the larger would-be instantons and
anti-instantons surviving these early cooling steps, all but $|Q|$ of
them will ultimately disappear due to the annihilation of
instanton--anti-instanton pairs.

\begin{figure}[t]
\begin{center}
\epsfig{figure=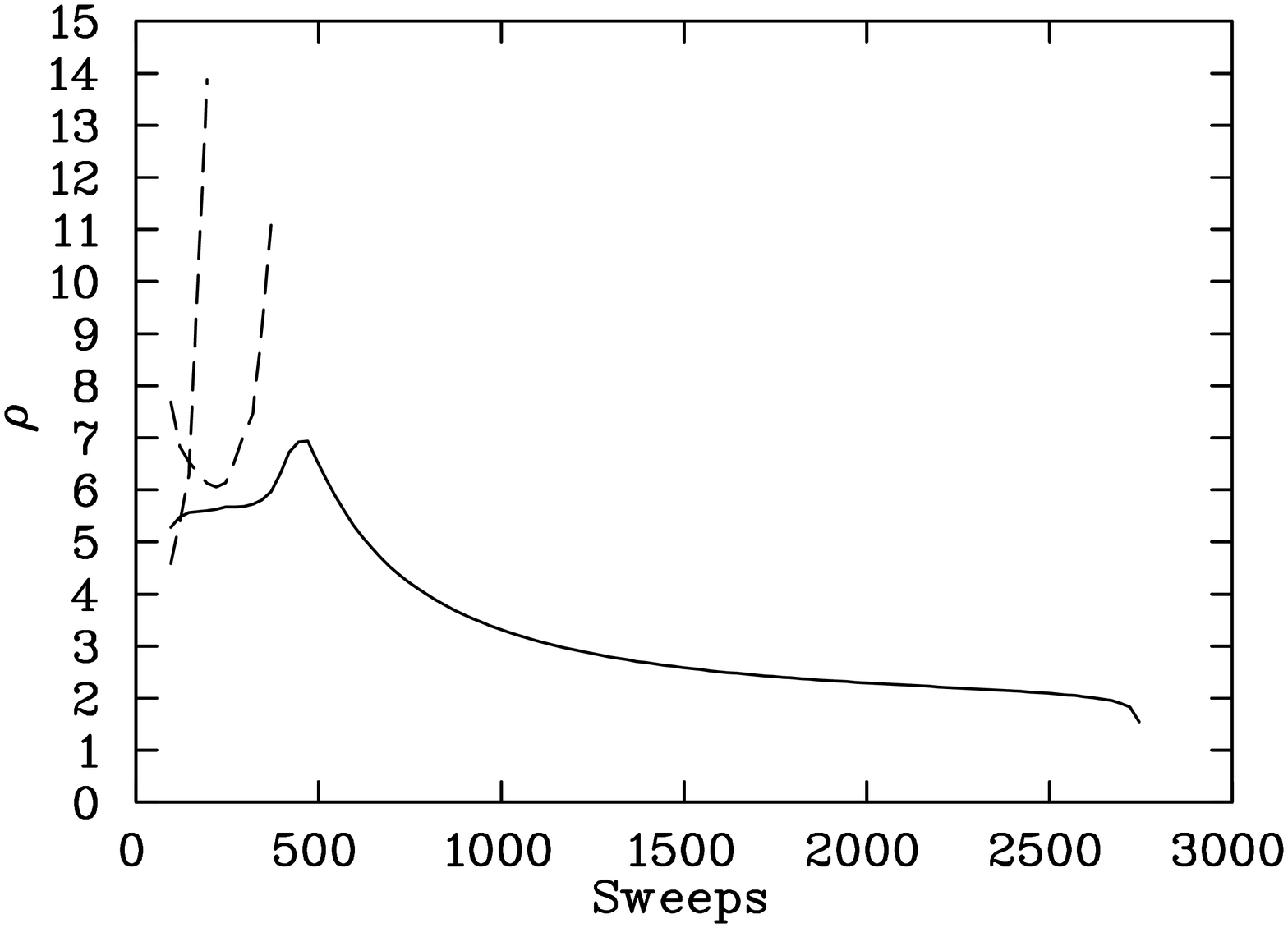,angle=90,width=\hsize}
\end{center}
\begin{center}
\epsfig{figure=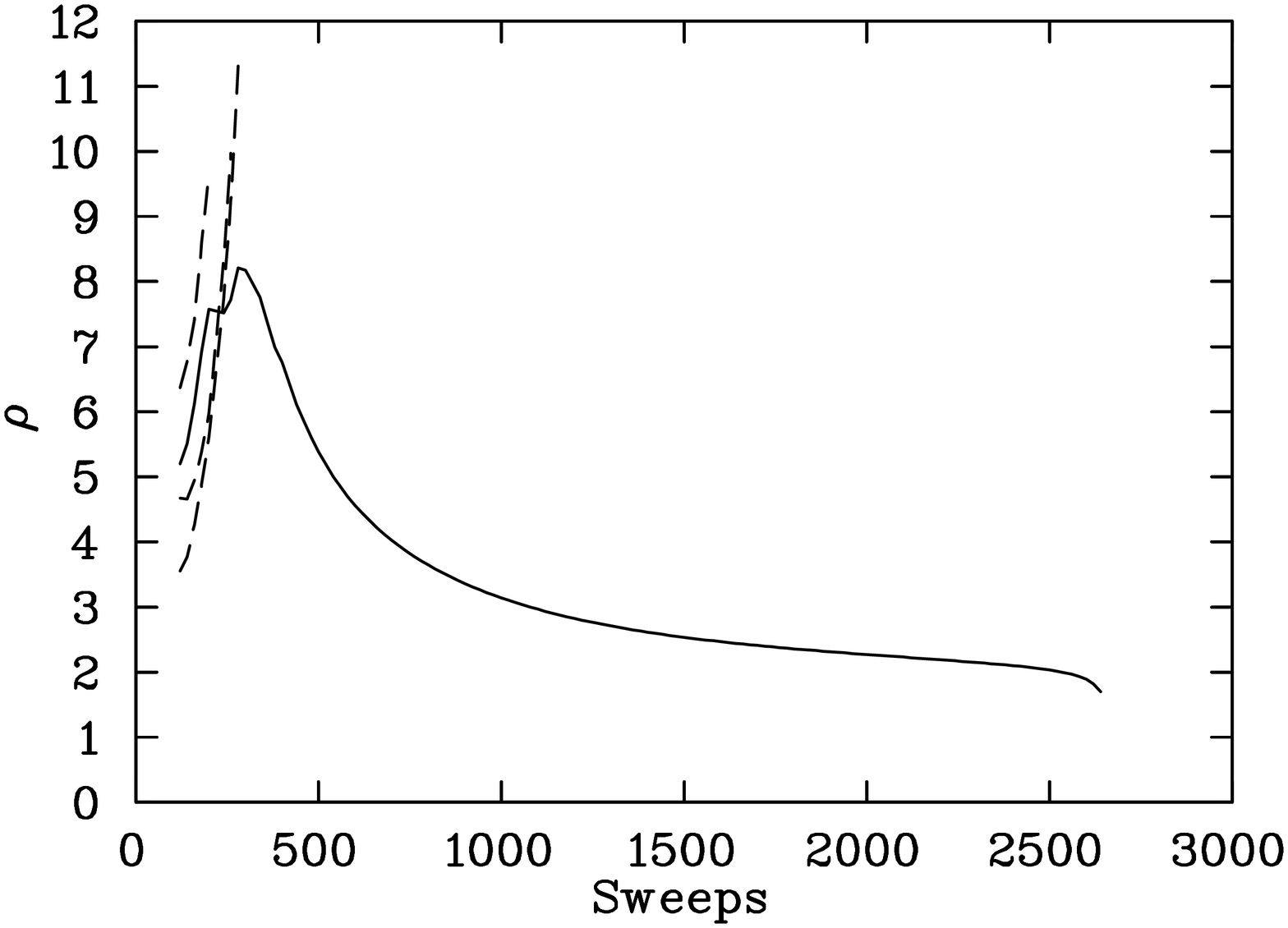,angle=90,width=\hsize}
\end{center}
\caption{Size of the single instanton ($\rho$ measured in lattice
units) in configuration 64 (upper figure) and configuration 11 (lower
figure).  An abrupt change in the size evolution occurs as other peaks
in the distribution disappear through the process of ``melting'' as
described in the text.  After sweep 2500 the instanton in each
configuration rapidly disappears and hence the associated size
parameter cannot be measured accurately.  Dashed lines represent the
size of transitory peaks.}
\label{fig:Qeq1cfg11&64rho}
\end{figure}

We use this algorithm to investigate the topological properties of
configurations 11 and 64 (corresponding to $|Q|=1$) and in
configurations 27 and 90 (corresponding to $|Q|=2$).  In
Fig.~\ref{fig:Qeq1cfg11&64rho} we illustrate how the instanton size,
$\rho$, varies with sweep number for configurations 64 and 11.  Three
initial instanton-like peaks are identified in configuration 64 and
four are identified in configuration 11.  The solid line in each of
these figures denotes the peak which survives long-term cooling and
becomes the approximately self-dual object in each configuration.  The
dashed lines denote the temporary peaks, which disappear under
cooling.  

In configuration 64 we see that all peaks rapidly grow in size
initially.  Both of the  temporary peaks are seen to expand in
size with the associated peak height decreasing until the peaks
``melt'' away.  These peaks are associated with the plateau in the
action observed in Fig.~\ref{fig:cfg64zout}.  Hence we see that
topological objects can disappear on the lattice by ``melting away'' 
where $\rho\to \infty$, which is a different mechanism from the 
behavior of ``falling through the lattice'' where $\rho\to 0$.

The remaining topological object reacts by slowing its growth.  Once
all of the other structures have disappeared, the surviving object
shrinks rapidly.  In minimizing its size relative to the finite volume
of the 4-torus, the object can best approximate the self-dual nature
of an instanton and temporarily evade the ultimate consequence of the
Nahm transform.  The rate of shrinkage is related to the size of the
object.  The shrinkage slowly continues until the instanton size is
approximately two lattice spacings, at which time the object is
smaller than the dislocation threshold of the cooling algorithm and is
rapidly suppressed.

Similar results are observed for configuration 11 (another $|Q|=1$
configuration), as shown in the lower plot of
Fig.~\ref{fig:Qeq1cfg11&64rho}.  Again all peaks rapidly grow in size
initially, but as each of the three temporary peaks expands and
``melts away'', the single instanton reacts by slowing its growth.
Once all of the other structures have disappeared, we see that the
surviving single (anti-)instanton shrinks rapidly due to its large
size.  The shrinkage continues until the dislocation threshold is
encountered, at which point it is rapidly suppressed.

The point of inflection seen in the evolution of the single-instanton
size around 1900 sweeps in Fig.~\ref{fig:Qeq1cfg11&64rho} signifies
the dislocation threshold of the S(3) improved cooling algorithm.  For
both configurations $\rho \simeq 2.2$ at the point of inflection. This
dislocation threshold is similar to $\rho_{\mathrm D} = 2.23$ for
$S(5)$ as determined in Ref.~\cite{deForcrand1997}.  

In the continuum infinite volume limit, instantons have no implicit
scale and can be any size with the same action $S_0$.  On a finite
4-torus in the continuum, the only relevant quantity is the relative
size of the instanton to the 4-torus size.  On the continuum 4-torus,
we therefore expect that in the limit where the size of the would-be
instanton vanishes with respect to the 4-torus size, it should able to
approach arbitrarily closely to self-duality.  Thus, under continued
cooling on the lattice a $|Q|=1$ configuration will approach the
self-dual limit by decreasing its size.  As it does so, discretization
errors will become increasingly large until the object encounters the
dislocation threshold, at which point the object will be removed by
the cooling algorithm. This is the ultimate fate of a $|Q|=1$
configuration under cooling.

In both $|Q|=1$ cases, the onset of the cresting in the topological
charge occurs when the size of the object is seven to eight lattice
spacings, well above the size of two lattice spacings where the
dislocation threshold of the cooling algorithm removes the objects.
Thus, at the onset of the cresting of $|Q|$ the objects are large
enough that the dislocation threshold of the cooling algorithm is
irrelevant. Since the cresting behavior appears to be a distinct
signal of the ultimate disappearance of the $|Q|=1$ objects, this
suggests that the dislocation threshold is not the primary reason for
the instability of the $|Q|=1$ configurations, even though it is the
reason for its final disappearance.

To make this even more concrete, we contrast these results for the
size dependence of the $|Q|=1$ objects under cooling, with the
corresponding results for the $|Q|=2$ configurations.

\begin{figure}[t]
\begin{center}
\epsfig{figure=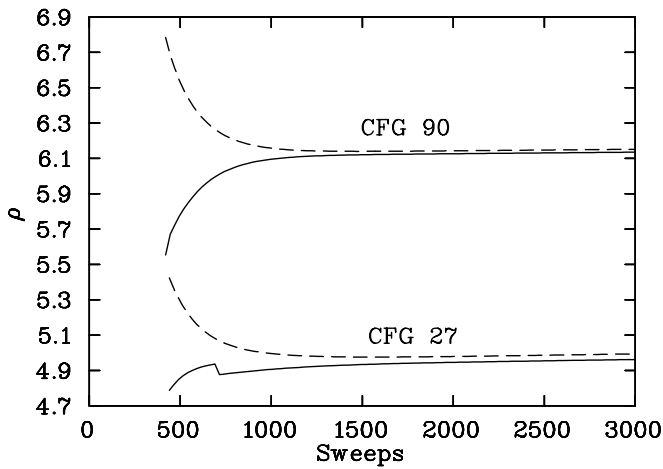,angle=90,width=\hsize}
\end{center}
\caption{Size evolution of the two instantons ($\rho$ measured in
lattice units) in configuration 27, and the two instantons in
configuration 90.  The solid and dashed lines represent the size of one
of the two instantons present in each configuration. The discontinuity
in the lower figure (configuration 27) is believed to be caused by a
sudden change in parameter values as the fitting program used to
calculate the single instanton size from our lattice data jumped from
one local $\chi^2$ minimum to another.}
\label{fig:Qeq2cfg90rho}
\end{figure}

The sizes of the two instantons in each of configurations 27 and 90
(which each have $|Q|=2$) are shown in Fig.~\ref{fig:Qeq2cfg90rho}.
First, note that in each case the configuration consists of two
identifiably separate topological objects, rather than one single
object of instanton number 2. For each configuration, the two
instantons drift apart under cooling, appearing to repel one
another. More interestingly, in each case, the size of one instanton
grows while the other shrinks, until they each settle (after roughly
1000 sweeps) to approximately the same size.  We do not know if this
behavior is necessary -- however, the fact that the same thing
happened in these two independent cases is suggestive.  Perhaps this
symmetry between the sizes of the cooled two-instanton configuration
is due to an image effect that balances two objects at equal sizes
roughly at ``opposite ends'' of the lattice. It would be interesting
to probe this question more deeply, both on the lattice and for the
continuum four-torus. For example, for $SU(2)$ ADHM instantons on
${\bf R}^4$ it is known that the behavior of $Q=2$ instantons depends
crucially on the relative $SU(2)$ orientation of the two constituent
instantons, which has important implications for the instanton size
distribution \cite{orientation}.

Another striking contrast between the size-dependence of the $|Q|=2$
and $|Q|=1$ configurations is that the $|Q| \simeq 1$ objects shrink
in size by a factor of approximately three as the cooling proceeds
over the course of two thousand sweeps, while in the $|Q| = 2$
configurations, each object changes size by no more than 10\%. This
indicates a significant level of stability throughout most of the
course of the cooling process. Since the $|Q|=2$ configurations
consist of two distinct topological objects, each of which may be
thought of as a single instanton, it is striking that for these
configurations the single instantons do not shrink in the same way
that the $|Q|=1$ would-be instantons do. This shows that the
distinction between the $|Q|=1$ and $|Q|=2$ configurations is really a
global effect, not a local one, and this is exactly what we would
expect from the Nahm transform corollary \cite{braam}.  So, this
difference in the size dependence under cooling is another signal (in
addition to the cresting behavior of $|Q|$ identified in the previous
section) of the instability of $|Q|=1$ configurations under cooling,
and once again we note that the difference can be seen already very
early in the cooling process, rather than just at the late time when
the $|Q|=1$ objects vanish due to discretization errors.

\subsection{Discretization Errors}
\label{sec:discussion}

\begin{figure}[t]
\begin{center}
\epsfig{figure=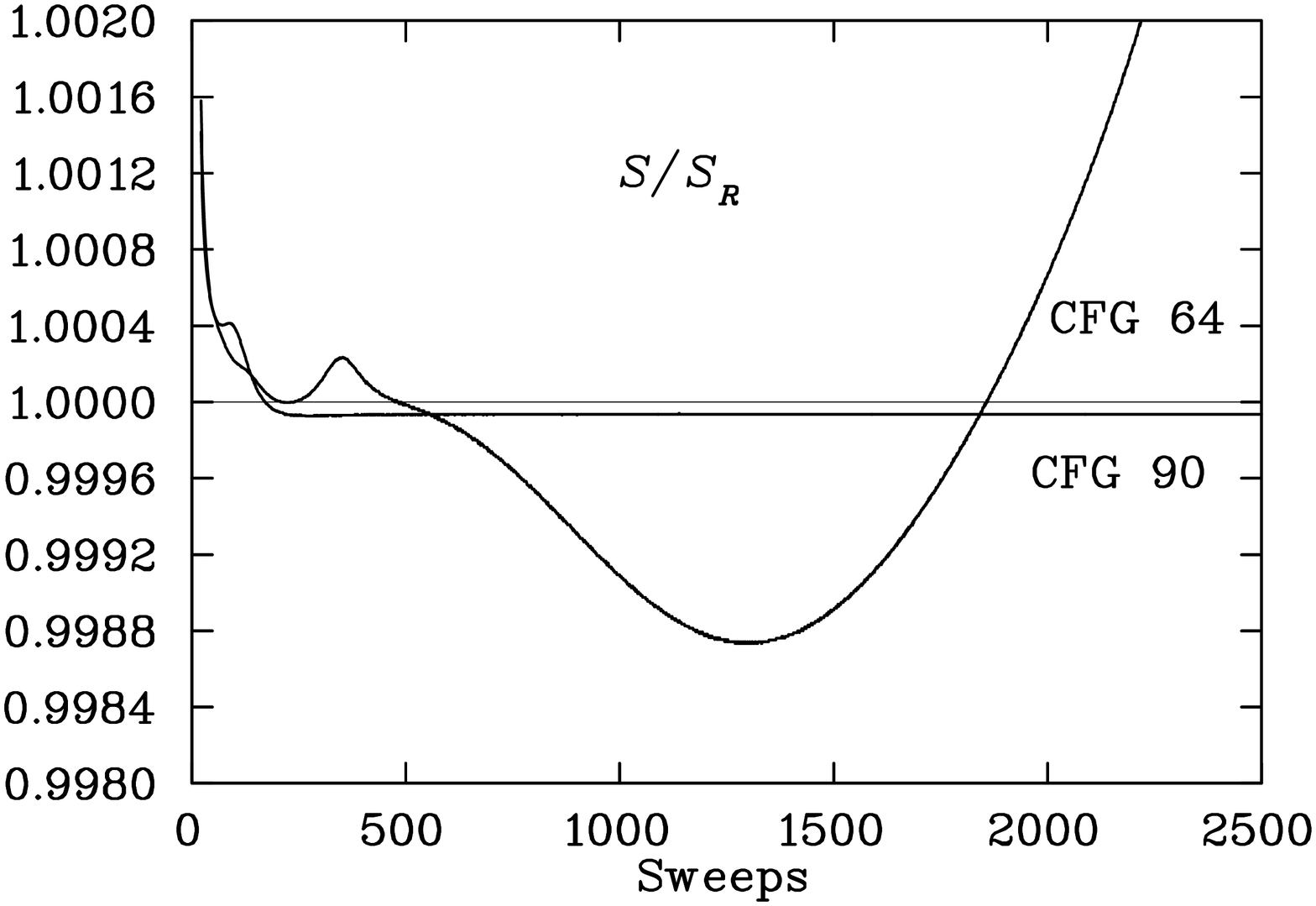,angle=90,width=\hsize}
\end{center}
\begin{center}
\epsfig{figure=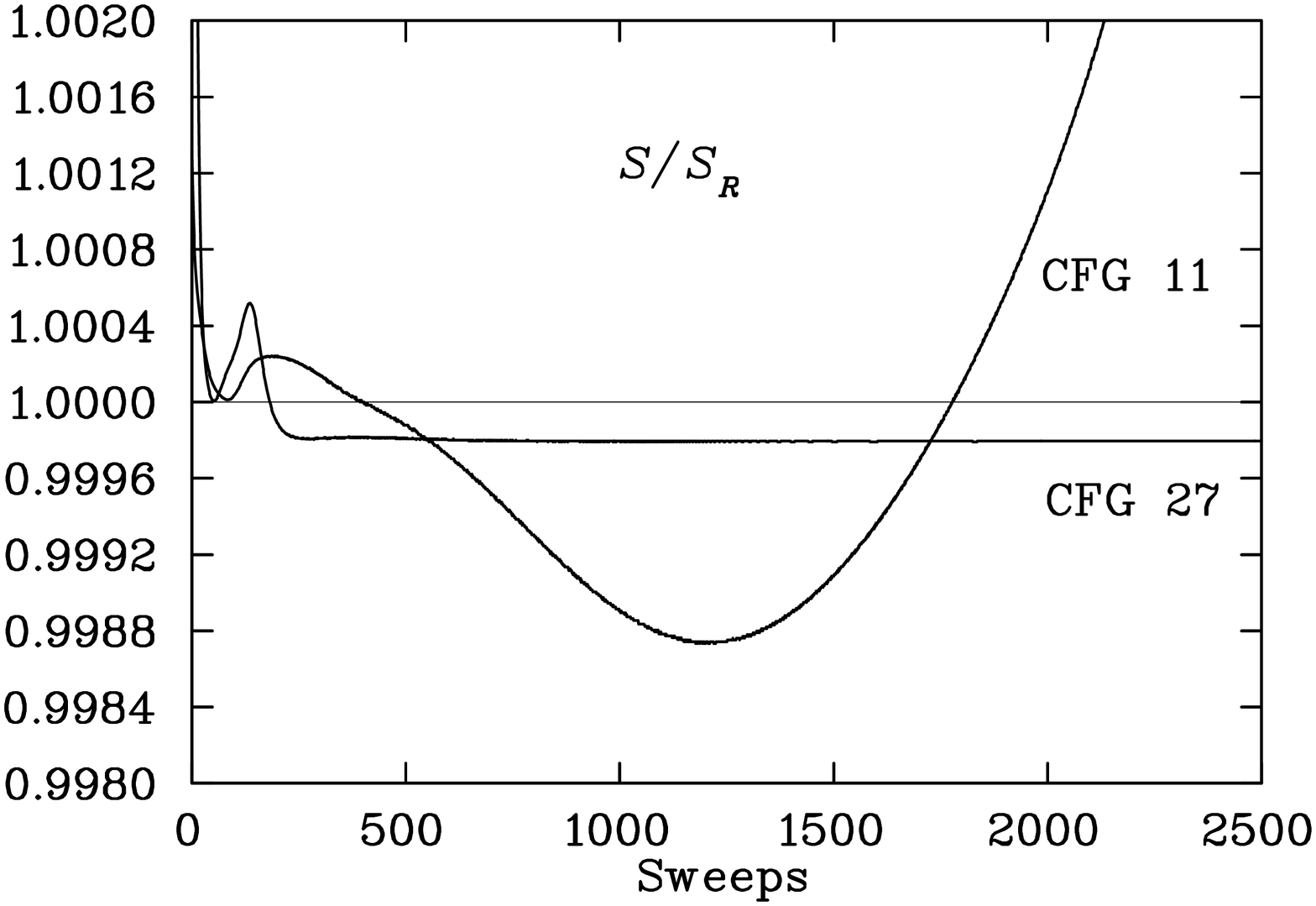,angle=90,width=\hsize}
\end{center}
\caption{Ratio of the cooling action to the reconstructed action for
configurations 64 and 90 (top) and 11 and 27 (bottom).  Configurations
11 and 64 are $|Q| \simeq 1$ configurations, whereas configurations 27
and 90 are $|Q| = 2$ configurations.  }
\label{fig:SonSRratios}
\end{figure}

In this last section we identify a third distinct signal of the
instability of the $|Q|=1$ configurations under cooling, and show that
it can also be seen early in the cooling process.  It was noted
already in Figs.~\ref{fig:cfg64zin} and \ref{fig:cfg11zin} that $S$
and $S_R$ diverge rapidly in the period prior to the collapse of the
$|Q| = 1$ configurations.  Actually, for the $|Q|=1$ configurations,
the deviation between these two actions appears much earlier, as is
shown in Fig.~\ref{fig:SonSRratios}.  In Fig.~\ref{fig:SonSRratios} we
present plots of the ratio of $S$ to $S_R$, for both $|Q| = 1$
configurations (11 and 64) and $|Q| = 2$ configurations (27 and 90).
Since the cooling action and reconstructed action are improved
differently, they are expected to have different ${\cO}(a^6)$
discretization errors.  The ratio of $S$ to $S_R$ therefore indicates
the relative scale of the ${\cO}(a^6)$ errors present.

Early in the cooling evolution, peaks in $S/S_R$ are seen to be
associated with the disappearance with topological structures as
illustrated in Fig.~\ref{fig:Qeq1cfg11&64rho}.  In the case of the
single-instanton configurations, the ratio of the cooling action to
reconstructed action, $S/S_R$, diverges from unity slowly during the
middle stages of the cooling process.  The point where this ratio
starts falling below 1 corresponds to the beginning of the cresting
behavior of $|Q|$ which was identified for the $|Q|=1$ configurations
in Section \ref{cresting} .  This corresponds to the early stage where
the difference between the size evolutions of the $|Q|=1$ and $|Q|=2$
configurations becomes clear.  Also, the position of the valley in the
ratio $S/S_R$ corresponds to the peak of the cresting of the
topological charge illustrated in Figs.~\ref{fig:cfg64zin} and
\ref{fig:cfg11zin}.

Finally, $S/S_R$ diverges rapidly as the object present in each
configuration falls through the lattice.  As the instanton shrinks,
the ${\cO}(a^6)$ errors overwhelm the improved action operator and
large errors allow the disappearance of the object. By contrast, 
in the two $|Q|=2$ configurations the ratio $S/S_R$ remains within 
0.02\% of unity indefinitely.

\section{Conclusions}
\label{sec:Conclusions}

In this paper we have compared the cooling of $|Q|=1$ and $|Q|=2$
configurations in $SU(3)$ gauge theory, using improved lattice actions
and precise monitoring of the sizes of the topological objects and of
the lattice discretization errors. In the continuum it is not possible
to have a self-dual $|Q|=1$ configuration on the untwisted four-torus
\cite{braam}, and the results from our lattice analysis suggests three
distinct signals, each of which occurs very early in the cooling
process, that the $|Q|=1$ configurations behave differently from the
$|Q|=2$ ones, and that they will ultimately shrink to below the
discretization threshold and then disappear much later in the cooling
process. These instability signals occur at a stage in the cooling
when the topological objects are still much larger than the lattice
discretization threshold. None of these signals is present for the
$|Q|=2$ configurations we studied, even though these consisted of two
isolated single instantons. This fact is a clear indication that these
$|Q|=1$ instability signals are reflecting the {\it global} properties
of the torus, which is the essence of the continuum torus result
\cite{braam}, rather than {\it local} effects such as those that
ultimately cause a very small $|Q|=1$ object to fall through the
lattice due to discretization errors.

The three instability signals we have found to be characteristic of
the $|Q|=1$ configurations are as follows. First, there is a cresting
behavior of the topological charge $|Q|$ away from its precise integer
value of 1, as shown in Figures 2 and 3.  This does not happen for the
$|Q|=2$ configurations, as shown in Figures 4 and 5. Second, for the
$|Q|=1$ configurations there is an initial swelling of the size of the
would-be instanton, but then a steady shrinkage begins at the same
point where the cresting in $|Q|$ is observed. This is illustrated in
Figure 6. Once again, this is completely different from the behavior
for the $|Q|=2$ configurations, which is shown in Figure 7. The third
signal is that for the $|Q|=1$ configurations the ratio $S/S_R$ of the
cooling action to the reconstructed action begins a steady fall below
one, before bottoming out and eventually diverging much later in the
cooling process, as shown in Figure 8. This deviation below a ratio of
1 begins at the same point, early in the cooling process, where the
cresting of $|Q|$ and the rapid shrinkage of the object begin. This is
also very different from the behavior of the $|Q|=2$ configurations,
already at this early stage where the topological objects are large.

\begin{figure}[t]
\begin{center}
\epsfig{figure=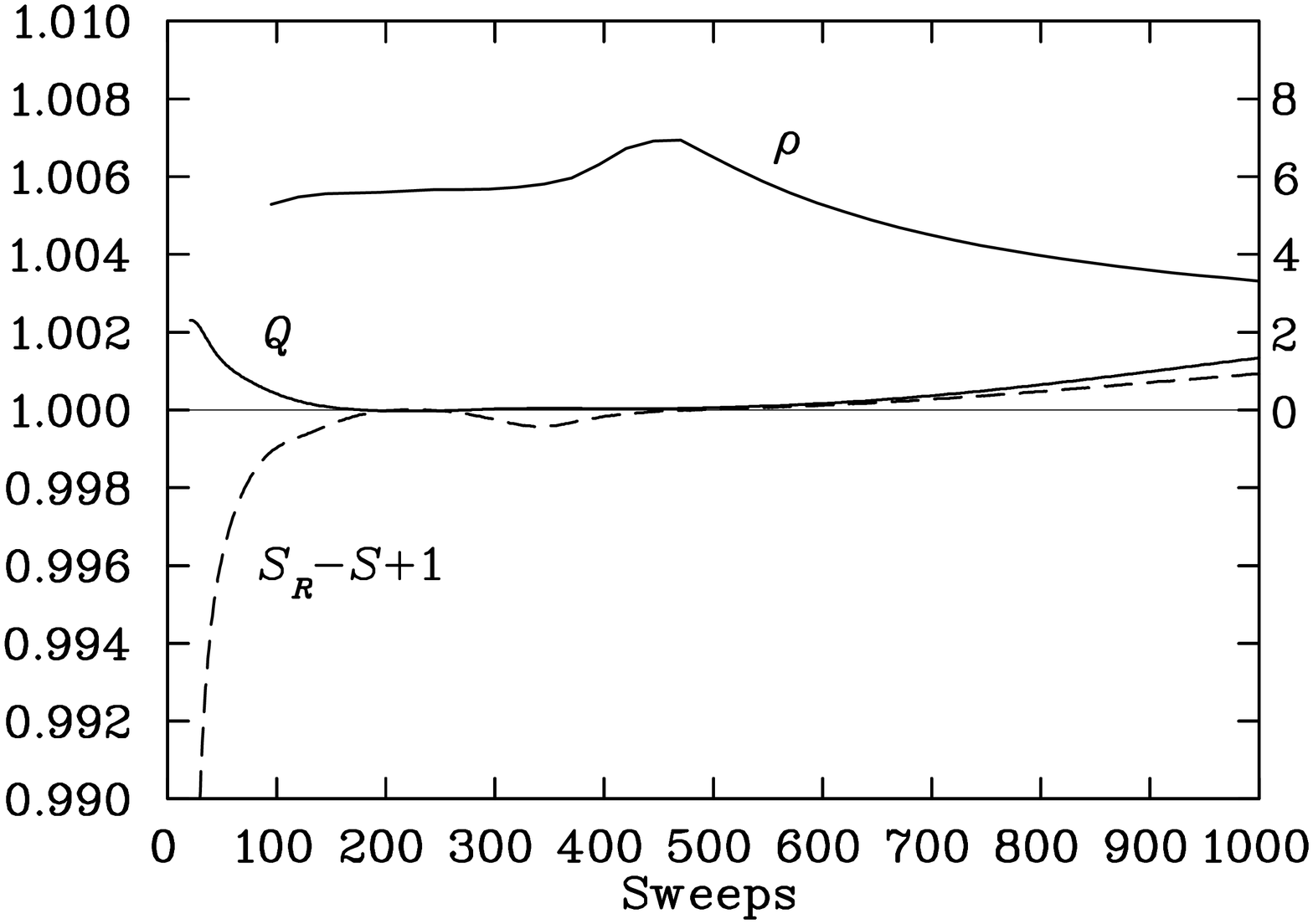,angle=90,width=\hsize}
\end{center}
\begin{center}
\epsfig{figure=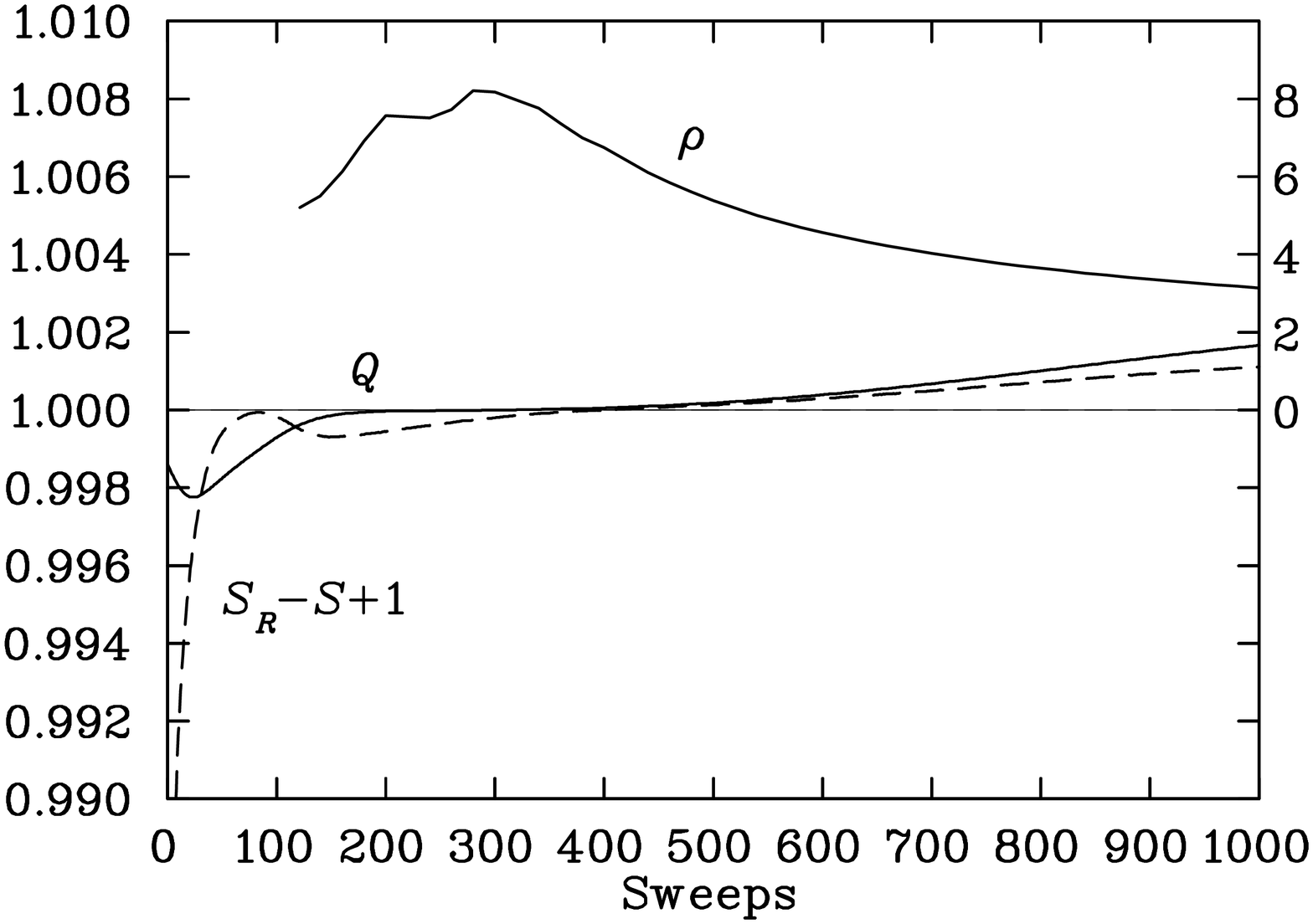,angle=90,width=\hsize}
\end{center}
\caption{The behavior of instanton peak size, $\rho$ (upper solid
  line, right scale), against topological charge (lower solid line, 
  left scale) and $S_R-S+1$ (dashed line, left scale), as a function
  of sweep number, for $|Q| \simeq 1$ configurations 64 (upper
  figure) and 11 (lower figure).  Notice that $S_R-S+1$ and $Q$ move
  away from integer values at the same time that the instanton starts
  to shrink.  Cooling is performed with the $S(3)$ action, and the
  topological charge is $Q(3)$. We have taken the value $S_R-S+1$
  rather than $S_R-S$ so that the action and charge curves can be
  plotted on the same precise scale.}
\label{fig:cfg64sizeandSQ}
\end{figure}

These results suggest the following picture of the instability, and
ultimate disappearance, of $|Q|=1$ configurations under cooling. The
first stage sees the rapid growth and disappearance of temporary peak
structures.  The second stage, whose onset is indicated by the three
signals listed in the previous paragraph, sees the shrinkage with
cooling of the single instanton. This type of shrinkage does not
happen for the separate constituents of the $|Q|=2$ configurations.
During this shrinkage stage the topological charge deviates slightly
from its integer value of 1.  The third and final stage is the rapid
disappearance of the single instantons due to dislocation errors at an
instanton size of approximately two lattice spacings.

To emphasize these points, we show in Fig.~\ref{fig:cfg64sizeandSQ}
the instanton size $\rho$, the topological charge $Q$, and the
difference $S_R-S+1$ between the reconstructed action and cooling
action for the $|Q|=1$ configurations 64 and 11 in the region where
the Nahm transform corollary manifests itself.  (The addition of one
unit to the difference serves to plot the action and charge curves on
the same precise scale.)  Since $S_R$ and $S$ have different
${\cO}(a^6)$ errors, their difference is an extremely effective probe
of the scale of discretization errors in the configuration.  As the
instanton changes from expanding behavior to shrinking behavior, the
topological charge and difference $S_R-S+1$ are both almost
identically equal to one, indicating minimal discretization errors at
the onset of the Nahm transform corollary.  This strongly suggests
that the sudden change in behavior of the would-be instanton is due to
the global nature of the field on the toroidal lattice, and not to
local discretization errors.

The $|Q|=2$ configurations behave very differently from the $|Q|=1$
configurations.  For each of the two $|Q|=2$ configurations we
studied, the configuration consisted of two isolated lumps that drift
apart under cooling and have the remarkable property that one lump
shrinks, while the other swells, until they reach the same size. It
would be interesting to know if this behavior is generic, due to some
torus periodicity-induced balance, and if there is any trace of such
behavior in the continuum. Unfortunately, this final question is a
difficult one, as there are no known nontrivial ({\it i.e.}\
inhomogeneous) analytic torus instantons.

We close by noting that all five criteria outlined in the
introduction have been met.  In particular:
\begin{enumerate}
\item The S(3) cooling action is very accurate with remaining errors
  positive, acting to stabilize topological structure.
\item The action and topological charge density of the $|Q| \simeq 1$
  configurations are seen to be distributed over large length scales
  of 7 to 8 lattice spacings, much larger than the dislocation
  threshold of the improved cooling algorithm at 2.2 lattice spacings.
\item Comparisons of the
  reconstructed action with the more traditionally constructed
  improved action have provided a  powerful way of
  investigating the scale of discretization errors in the
  configurations under investigation.
\item $|Q|=1$ configurations have been shown to be stable until the
  action $S$ drops below $3\, S_0$ and approaches the self-dual limit
  of $S_0 = 8 \pi^2 / g^2$.  Extra peaks in the action density
  disappear as $S/S_0$ drops from 3 toward 1.
\item Finally, the action and topological charge distributions are
  seen to approach the classical instanton form with only a single
  peak appearing and the instanton scale parameter $\xi \to 1$ as
  cooling proceeds.
\end{enumerate}

\begin{acknowledgments}
The calculations reported here were carried out on the Orion
supercomputer at the Australian National Computing Facility for
Lattice Gauge Theory (NCFLGT) at the University of Adelaide.  GD
thanks the CSSM at Adelaide for hospitality while this work was begun,
and the U.S. DOE for support through the grant DE-FG02-92ER40716.
Financial support from the Australian Research Council is gratefully
acknowledged.
\end{acknowledgments}


\begin{thebibliography}{20}
\bibitem{correction} Here we correct a typographical error in
Ref.~\cite{Bilson-Thompson:2002jk} where the normalization was
incorrectly stated as 1/2 as opposed to 1/12.
\end{thebibliography}
\end{document}